\documentclass[twocolumn,preprintnumbers,amsmath,amssymb]{revtex4}
\usepackage{graphicx}

\usepackage{amsmath}
\usepackage{amssymb}
\usepackage{amsthm}
\usepackage{amsfonts}
\usepackage{enumerate}

\usepackage{centernot}
\usepackage{tikz}

\usepackage{subfigure}

\begin{document}

\newcommand{\bq}{{\bf q}}

\title{Topological superconductor from superconducting topological surface states and fault-tolerant quantum computing}

  \author{Xi Luo$^1$}
  \thanks{These two authors contribute equally.}

 \author{Yu-Ge Chen$^{2,3,4}$}
 \thanks{These two authors contribute equally.}

 \author{Ziqiang Wang$^5$}
 \thanks{Correspondence to: wangzi@bc.edu}
 \author{ Yue Yu$^{2,3,4}$}
 \thanks{Correspondence to: yuyue@fudan.edu.cn}

 \affiliation {
 	1. College of Science, University of Shanghai for Science and Technology, Shanghai 200093, PR China\\
 	2. State Key Laboratory of Surface Physics, Fudan University, Shanghai 200433,
 	China\\
 	3.Center for Field
 	Theory and Particle Physics, Department of Physics, Fudan University, Shanghai 200433,
 	China\\
 	4. Collaborative Innovation Center of Advanced Microstructures, Nanjing 210093, China\\
 	5. Department of Physics, Boston College, Chestnut Hill, MA 02467, USA
 }

\date{\today}

\begin{abstract}
	{\bf The chiral $p$-wave superconductor/superfluid in two dimensions (2D) is the simplest and most robust system for topological quantum computation \cite{K1,TQCR} . Candidates for such topological superconductors/superfluids in nature are very rare. A widely believed chiral p-wave superfluid is the Moore-Read state in the $\nu=\frac{5}2$ fractional quantum Hall effect\cite{MR,RG}, although experimental evidence are not yet conclusive \cite{will}. Experimental realizations of chiral $p$-wave superconductors using quantum anomalous Hall insulator-superconductor hybrid structures have been controversial \cite{science,newscience}. Here we report a new mechanism for realizing 2D chiral $p$-wave superconductors on the surface of 3D $s$-wave superconductors that have a topological band structure and support superconducting topological surface states (SC-TSS), such as the iron-based superconductor Fe(Te,Se) \cite{arpes-fts}. We find that tunneling and pairing between the SC-TSS on the top and bottom surfaces in a thin film or between two opposing surfaces of two such superconductors can produce an emergent 2D time-reversal symmetry breaking chiral topological superconductor. The topologically protected anyonic vortices with Majorana zero modes as well as the chiral Majorana fermion edge modes ($\chi$MEMs)  can be used as a platform for more advantageous non-abelian braiding operations.
We propose a novel device for the CNOT gate with six $\chi$MEMs, which
paves the way for fault-tolerant universal quantum computing.
 }
\end{abstract}

\maketitle
The electronic structure of a large class of the crystalline materials is characterized by a topological $Z_2$ invariant.
Three dimensional (3D) strong topological insulators have a nontrivial $Z_2$ invariant band structure and support helical Dirac fermion topological surface states (TSS) protected by time-reversal symmetry. Several iron-based superconductors, highlighted by Fe(Te,Se), have emerged recently as charge transfer metals \cite{zhoukotliarwang} with nontrivial topological $Z_2$ invariant band structures originating from the $p$-$d$ band inversion and spin-orbit coupling (SOC) \cite{zjwang,jphu,gangxu}. In the normal state, these Fe-chalcogenides and pnictides have a small Fermi energy and can be considered as topological metals with lightly doped TSS, which have been observed by spin-polarized angle-resolved photoemission spectroscopy (ARPES) \cite{arpes-fts,pengzhang-natphys}. In the fully gapped superconducting (SC) state below $T_c$, ARPES shows that the TSS develop a SC gap, induced by bulk superconductivity, comparable to the bulk gap \cite{arpes-fts}. These remarkably discoveries suggest that superconductors bearing a topological nontrivial $Z_2$ invariant electronic structure can support a new form of quantum matter on their surfaces -- the SC topological surface states (SC-TSS).

\begin{figure}
	\includegraphics[width=0.4\textwidth]{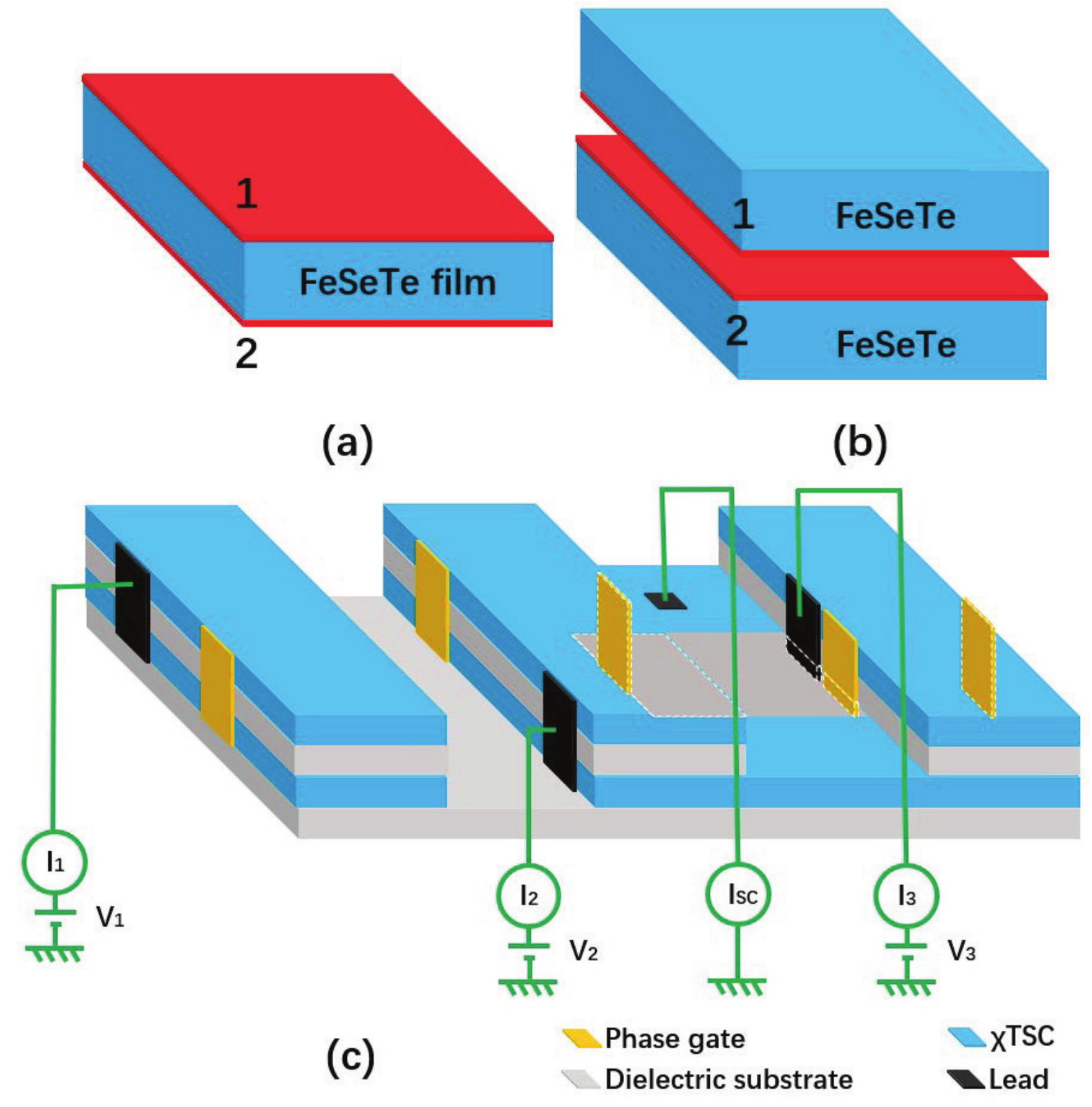}
	\caption{(color online)
		{The coupled SC-TSS illustrated with FeTeSe. (a) Thin film with top ($1$) and bottom ($2$) SC-TSS. (b) Opposing surfaces ($1$ and $2$) of two superconductors with SC-TSS. (c) Schematics of a device for the CNOT gate.
}
		\label{figX}	}
\end{figure}

The SC-TSS has provided a new platform for finding Majorana zero modes (MZMs) since the $\pi$-Berry phase of the Dirac fermions makes the vortex core states \cite{cdgm} of the SC-TSS carry integer total angular-momentum quantum numbers and naturally support the zero energy mode \cite{qav}. Indeed, the SC-TSS is a single-material realization of the Fu-Kane proposal for inducing superconductivity in the TSS of a strong topological insulator by proximity effect to an s-wave superconductor \cite{fukane}.
Candidate MZMs have been observed as zero-energy bound states inside external magnetic field induced vortices in FeTe$_{0.55}$Se$_{0.45}$ \cite{dinggao2018,hanaguri}, (Li$_{0.84}$Fe$_{0.16}$)OHFeSe \cite{donglai-prx}, CaKFe$_4$As$_4$ \cite{CaKFe4As4}, and quantum anomalous vortices \cite{qav} nucleated at the interstitial and adatom magnetic Fe sites in FeTe$_{0.55}$Se$_{0.45}$ \cite{yin-natphys2015,qav-mzm2020} and LiFeAs \cite{yin2019} without applying an external magnetic field. We report in this article that the SC-TSS can be used to produce an emergent nonmagnetic 2D time-reversal symmetry breaking chiral topological superconductor ($\chi$TSC).
The necessary ingredient is the
coupling of the SC-TSS on the top and bottom surfaces in a thin film (Fig.~\ref{figX}(a)) or between two opposing surfaces of two superconductors (Fig.~\ref{figX}(b)). Such $\chi$TSCs provide topologically protected anyonic vortices with MZMs as well as chiral Majorana fermion edge modes ($\chi$MEMs) for nonabelian braiding operations. As an example, a CNOT gate device, to be discussed in detail later, is shown in Fig.~\ref{figX}(c) using thin films (blue) hosting six $\chi$MEMs for universal quantum computing.

In the Nambu basis, the electron operator $\Psi_N=(\psi_{1\uparrow {\bf q}},\psi_{1\downarrow{\bf {q}}},\psi_{2 \uparrow{\bf {q}}},\psi_{2 \downarrow{\bf {q}}},-\psi^{\dag}_{1\downarrow {\bf {-q}}},\psi^{\dag}_{1\uparrow {\bf {-q}}},\psi^{\dag}_{2 \downarrow{\bf {-q}}},-\psi^{\dag}_{2\uparrow {\bf {-q}}})^T$, where $1$ and $2$ label the surface and $\uparrow$ and $\downarrow$ the pseudospin states in the presence of SOC. The system is described by
$
H={1\over2}\sum_{\bf q}\Psi_N^\dagger h({\bf q}) \Psi_N,
$
with the $8\times8$ Bogoliubov-de Gennes (BdG) Hamiltonian matrix
\begin{eqnarray}
h(\bq)&=&  h_0(\bq)-t\sigma_0\chi_x\tau_0 + h_{12}^\Delta
\label{h}\\
h_0(\bq)&=&v_F{\bf q}\cdot {\boldsymbol\sigma}\chi_z\tau_z-\mu\sigma_0\chi_0\tau_z+\Delta\sigma_0\chi_0\tau_x, \label{bdg}
\end{eqnarray}
where $(\boldsymbol{\sigma},\sigma_0)$ $({\boldsymbol\chi},\chi_0)$, and $(\boldsymbol{\tau},\tau_0)$ are Pauli and unit matrices acting in the pseudospin, surface, and particle-hole sectors, respectively.
$h_0(\bq)$
in Eq. (\ref{bdg})
describes the uncoupled SC-TSS, where $v_F$ is the Fermi velocity to be set to unity,
${\bf q}$ is the 2D momentum $(q_x, q_y)$, $\mu$ is the chemical potential, and the real $\Delta$ is the intra-surface $s$-wave pairing strength.
$h_0$ has inversion and particle-hole symmetries, namely, $\mathcal{P}h_{0}(-{\bf q})\mathcal{P}^{-1}=h_0(\bf q)$ and $\mathcal{C}h_0(-{\bf q})\mathcal{C}^{-1}=-h_0(\bf q)$ with $\mathcal{P}=\sigma_0\chi_x\tau_z$and $\mathcal{C}=\sigma_x\chi_y\tau_y\mathcal{K}$, respectively, where $\mathcal{K}$ is the complex conjugation operator.
The time-reversal (TR) operation under
$\mathcal{T}=i\sigma_y\chi_0\tau_0\mathcal{K}$ is also a symmetry of $h_0$ since $\mathcal{T}h_0(-{\bf q})\mathcal{T}^{-1}=h_0({\bf q})$. Clearly, $\mathcal{P}^2=1$, {$\mathcal{C}^2=1$}, and $\mathcal{T}^2=-1$. 

The couplings between the two surfaces are described by the single-particle tunneling $t$ and inter-surface pairing $h_{12}^\Delta$ in Eq.~(\ref{h}). We consider both spin singlet (surface triplet) and triplet (surface singlet) local pairing with real amplitudes,
\begin{eqnarray}
H_{12}^{\Delta_s}&=&\sum_{\bf q}\Delta_s(\psi_{1\uparrow{\bf q}}^{\dag}\psi_{2\downarrow{\bf -q}}^{\dag}-\psi_{1\downarrow{\bf q}}^{\dag}\psi_{2\uparrow{\bf -q}}^{\dag})+h.c.,\nonumber\\
	H_{12}^{\Delta_t}&=&\sum_{\bf q}\Delta_t(\psi_{1\uparrow{\bf q}}^{\dag}\psi_{2\downarrow{\bf -q}}^{\dag}+\psi_{1\downarrow{\bf q}}^{\dag}\psi_{2\uparrow{\bf -q}}^{\dag})+h.c.,\label{s-wave}\\ H_{12}^{\Delta_t^\sigma}&=&\nonumber\sum_{\bf q} \Delta_t^\sigma \psi_{1\sigma{\bf q}}^{\dag}\psi_{2\sigma{\bf -q}}^{\dag}+h.c.
\end{eqnarray}
The spin-singlet pairing $H_{12}^{\Delta_s}$ is TR invariant (even under ${\cal T}$). In contrast, the one-dimensional representation of the spin-triplet pairing, $H_{12}^{\Delta_t}$ \emph{breaks the TR symmetry} (odd under ${\cal T}$).
Similarly, for the two-dimensional equal-spin triplet pairing, the independent basis functions $H_{12}^{\Delta_t^\uparrow}\pm H_{12}^{\Delta_t^\downarrow}$ are odd and even under ${\cal T}$ respectively \cite{note1}.
These inter-surface pairings of the SC-TSS bear a formal analogy to the interorbital pairing considered for bulk Cu$_x$Bi$_2$Se$_3$ \cite{FB,FL}, where purely imaginary spin-triplet $\Delta_t$ pairing was argued to produce a class DIII TR invariant TSC in 3D \cite{ram,ram1,ktheory}.
For simplicity, we will not consider the equal-spin pairing channels further, since they do not introduce new physics. Thus we have,
\begin{eqnarray}
h_{12}^\Delta=-\Delta_s\sigma_0\chi_y\tau_y +\Delta_t\sigma_z\chi_x\tau_x,
\label{h12pairing}
\end{eqnarray}
in $h(\bq)$ of Eq.~(\ref{h}). The coupling of the SC-TSS by the tunneling and pairing terms ensures the breaking of TR symmetry by a novel nonmagnetic mechanism (see the SI \cite{SI} for more detailed symmetry analysis) and the emergence of a chiral superconductor.

\begin{figure}
\begin{minipage}{0.23\textwidth}
	\centerline{\includegraphics[width=1\textwidth]{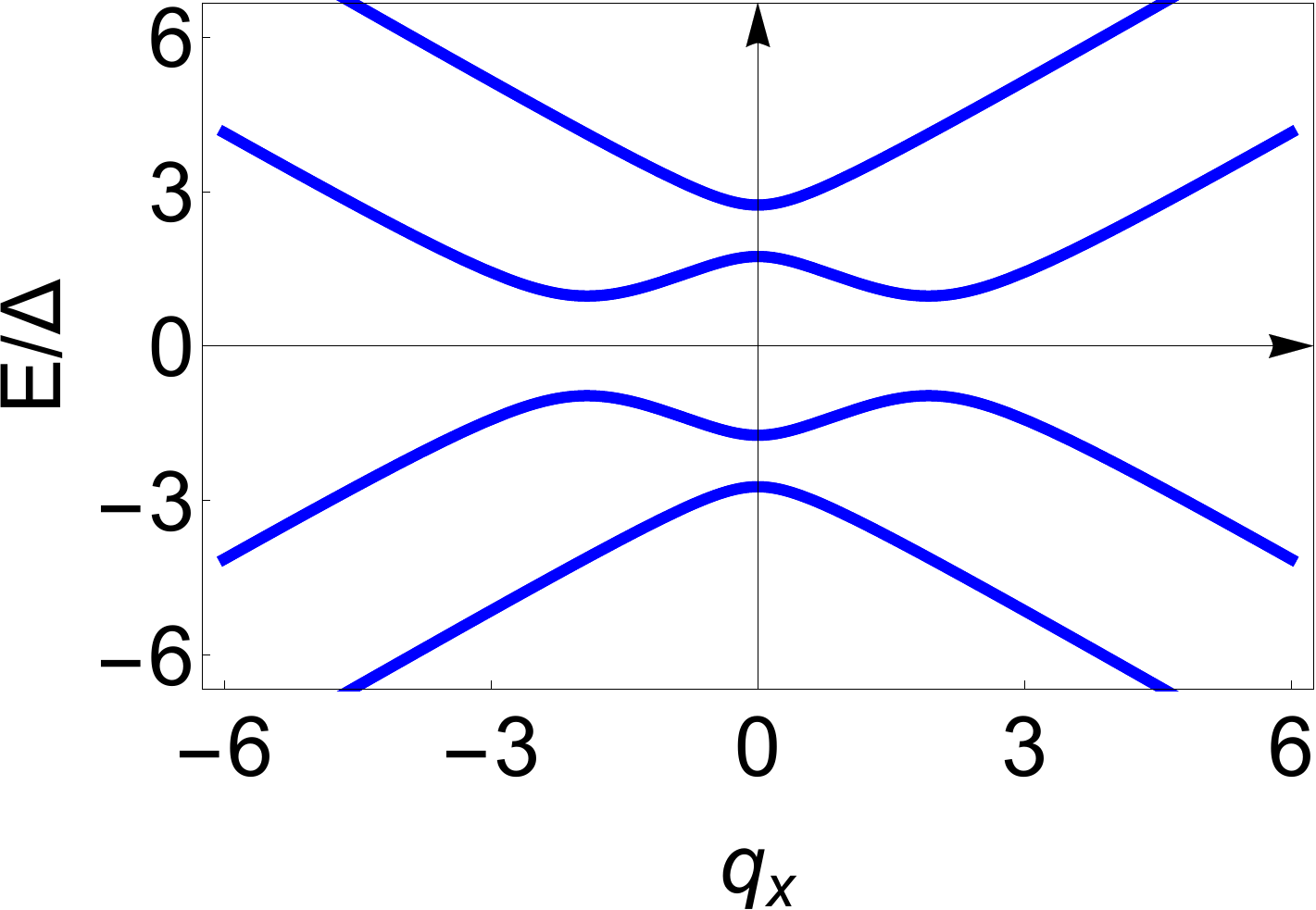}}
	\centerline{(a)}
\end{minipage}
\begin{minipage}{0.23\textwidth}
	\centerline{\includegraphics[width=1\textwidth]{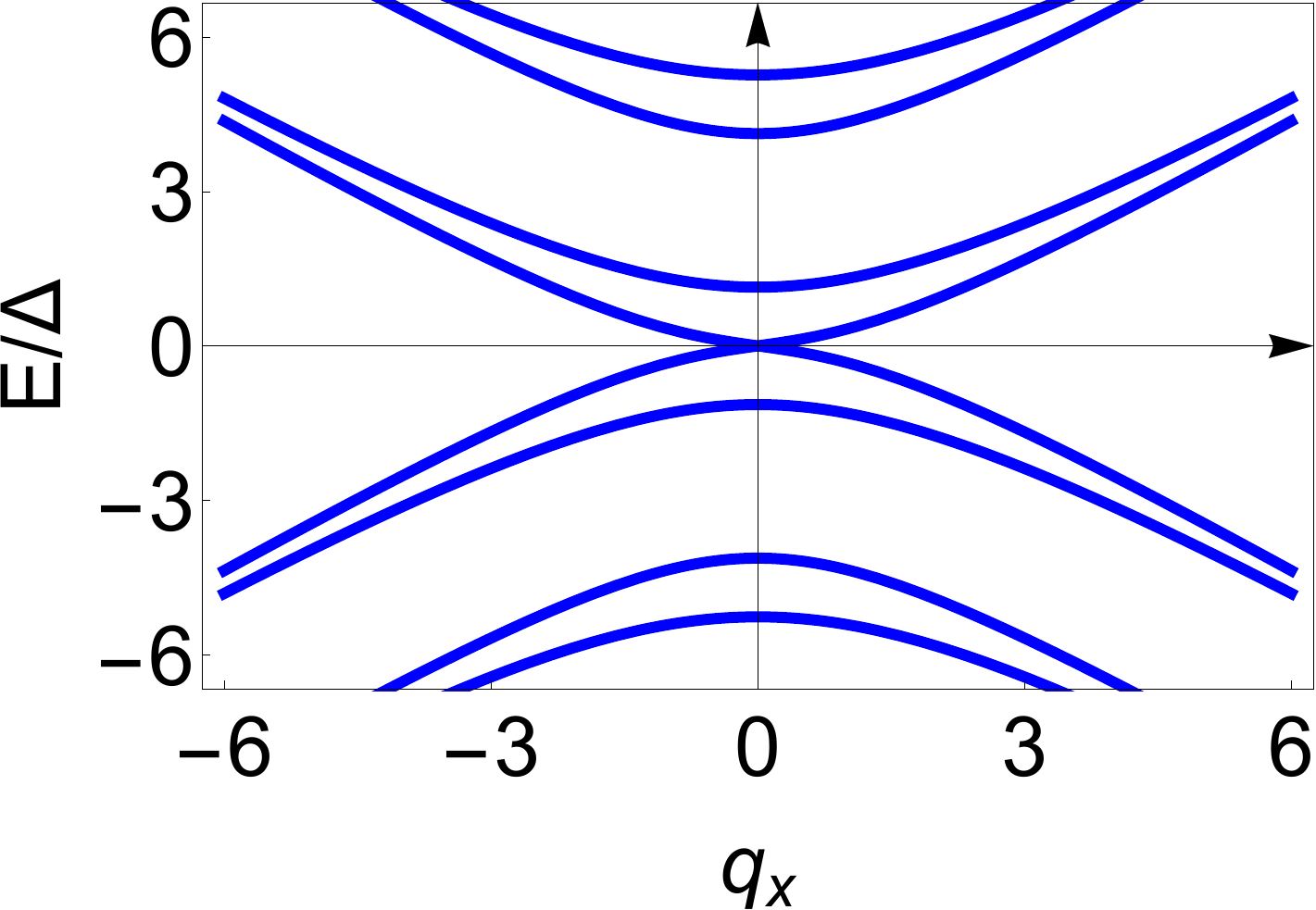}}
	\centerline{(b)}
\end{minipage}
	 	\begin{minipage}{0.23\textwidth}
	 		\centerline{\includegraphics[width=1\textwidth]{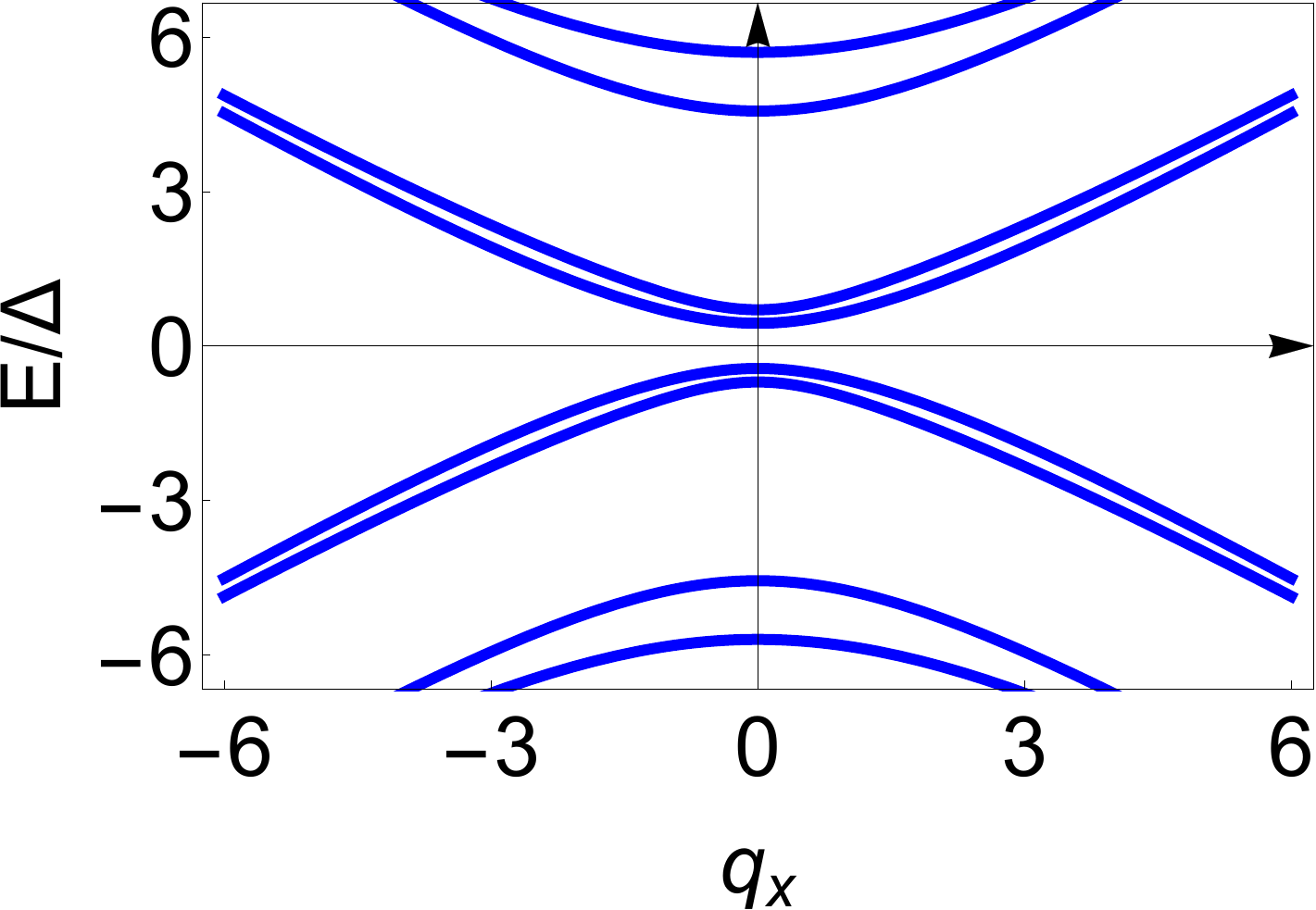}}
	 		\centerline{(c)}
	 	\end{minipage}
	 	\begin{minipage}{0.23\textwidth}
	 		\centerline{\includegraphics[width=1\textwidth]{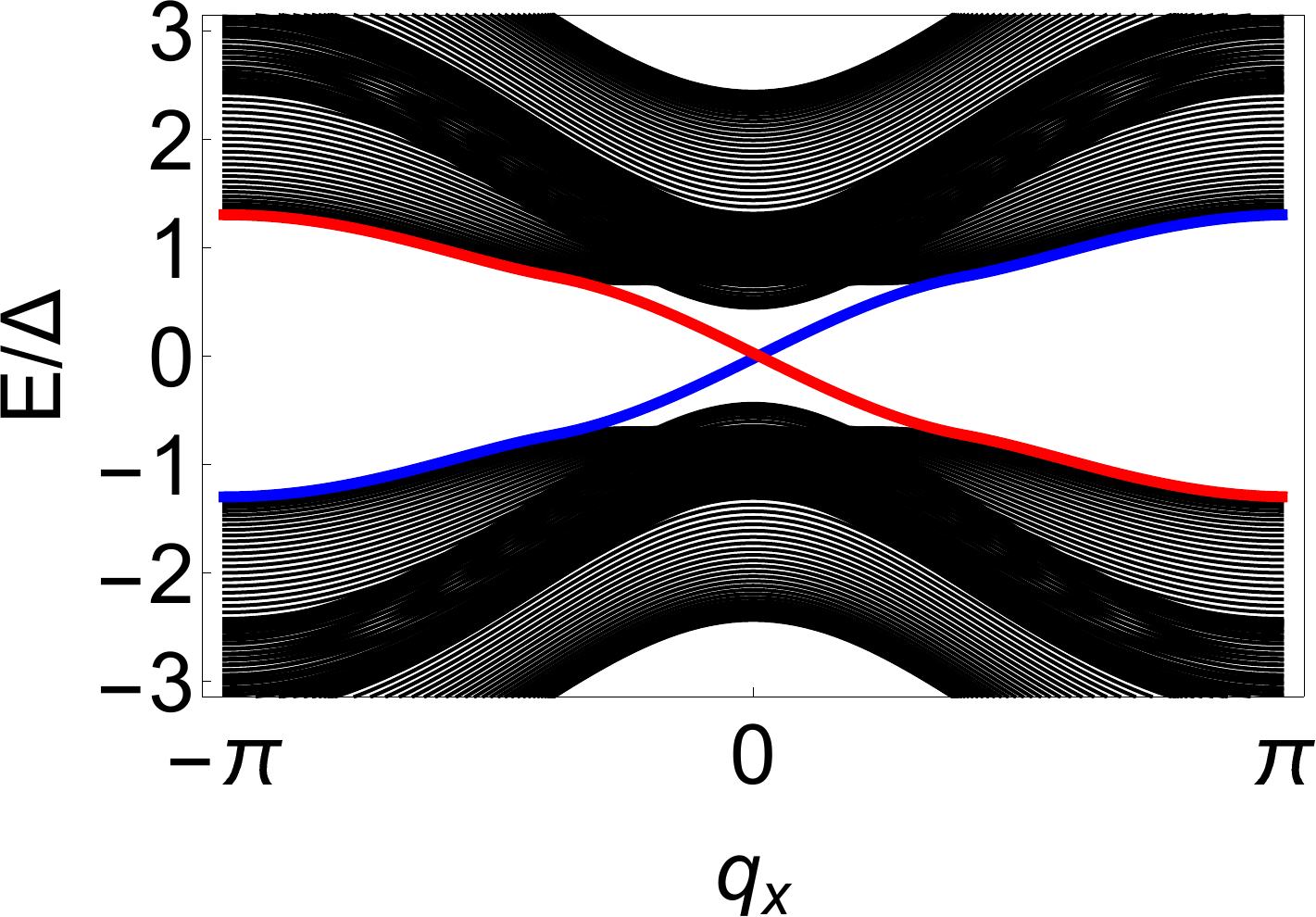}}
	 		\centerline{(d)}
	 	\end{minipage}
	\caption{(color online)
		Energy spectrum evolution of BdG Hamiltonian $h(\bq)$ in Eq.~(\ref{h}), plotted in the $q_y=0$ plane for $\Delta=1$, $\mu=2$, and $\Delta_s=0$.
(a) $t=0.5$, $\Delta_t=0$. (b) $t=2.06$, $\Delta_t=1.5$. (c) $t=2.5$, $\Delta_t=1.5$.
{(d) Energy spectrum of the lattice model with same parameters as in (c) under open boundary condition in $y$-direction (50 sites). The red and blue curves are the gapless edge states of a single $\chi$MEM. }
		\label{fig1}	}
\end{figure}

To illustrate the intrinsic ${\cal T}$-breaking phase induced by $\Delta_t$, we obtain the evolution of the energy spectrum (Fig.~\ref{fig1}) of the BdG Hamiltonian for $\Delta_s=0$.
The effect of a nonzero $\Delta_s$ will be discussed later. The values of $\mu$ and $\Delta$ are chosen in the range of the known experimental data for Fe$_{1+y}$Se$_x$Te$_{1-x}$ with $\Delta/\mu=0.5$ and {$\Delta\sim2$meV} \cite{dinggao2018,d-mu} which is set to unity.
We begin with the case $\Delta_t=0$ shown in Fig.~\ref{fig1}a, where the intra-surface $s$-wave pairing $\Delta$ opens a SC energy gap and the inter-surface tunneling $t$-term further gaps out the two Dirac points separated by $2\mu$. Each band in Fig.~\ref{fig1}a is doubly degenerate. Turning on ${\Delta_t}\neq0$, the degeneracy is lifted due to the broken TR symmetry. Increasing $t$ or {$\Delta_t$} leads to a gap closing at $\Gamma$ point ({Fig.~\ref{fig1}b}). The gap then reopens with a band inversion ({Fig.~\ref{fig1}{c}}) triggering a topological phase transition into the emergent $\chi$TSC that supports a single gapless Majorana edge mode localized at the open boundaries ({Fig.~\ref{fig1}{d}}). This is confirmed by a lattice model study \cite{SI}, and is consistent with a bulk Chern number $N=1$.
Increasing the tunneling $t$ can cause the gap to close again and the system to return to a topologically trivial superconductor {(not shown)}.

\begin{figure}
	\begin{minipage}[b]{0.23\textwidth}
	\centering
	\subfigure[]{
		\includegraphics[width=1\textwidth]{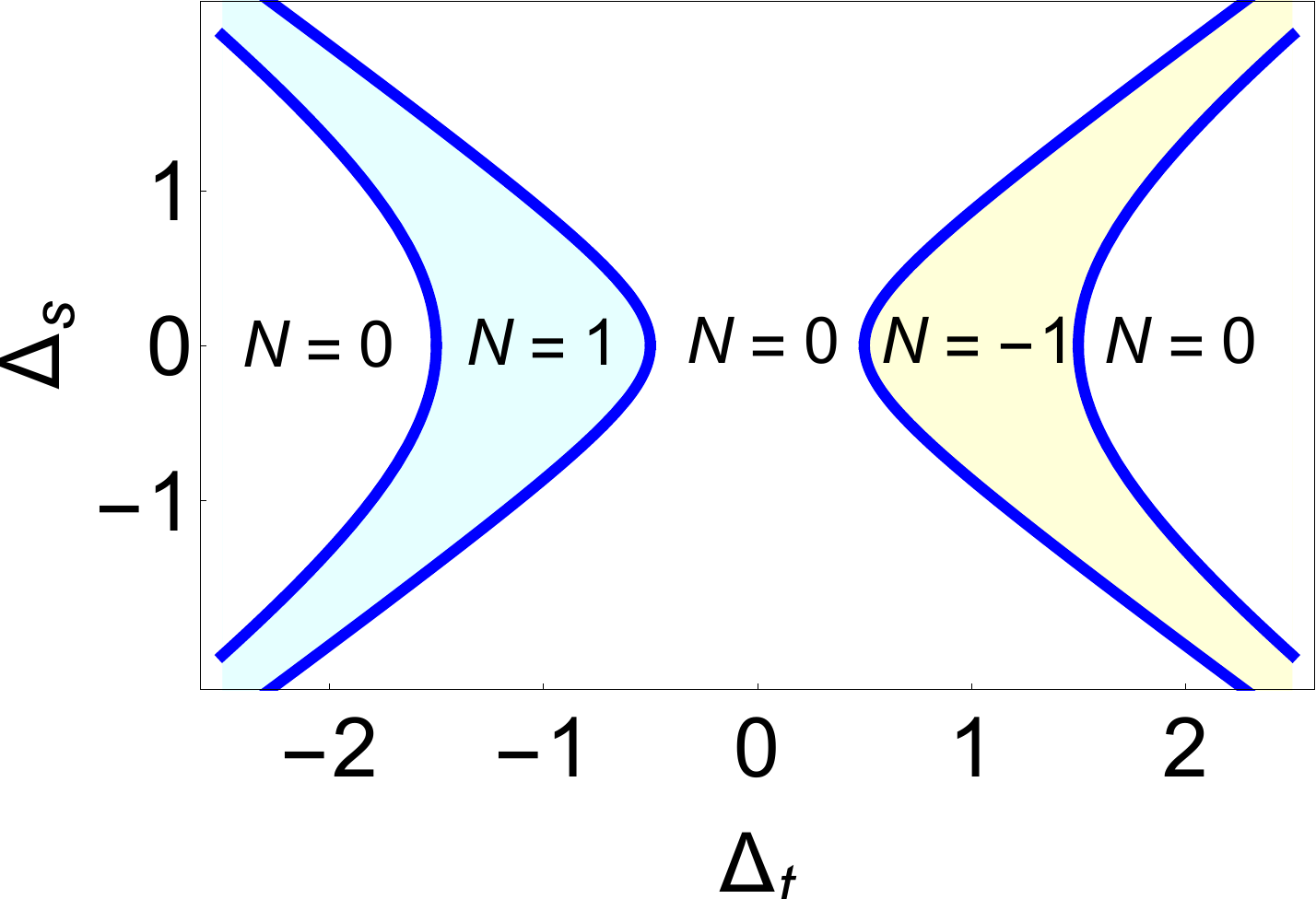}
	}	
	
	\subfigure[]{
		\includegraphics[width=1\textwidth]{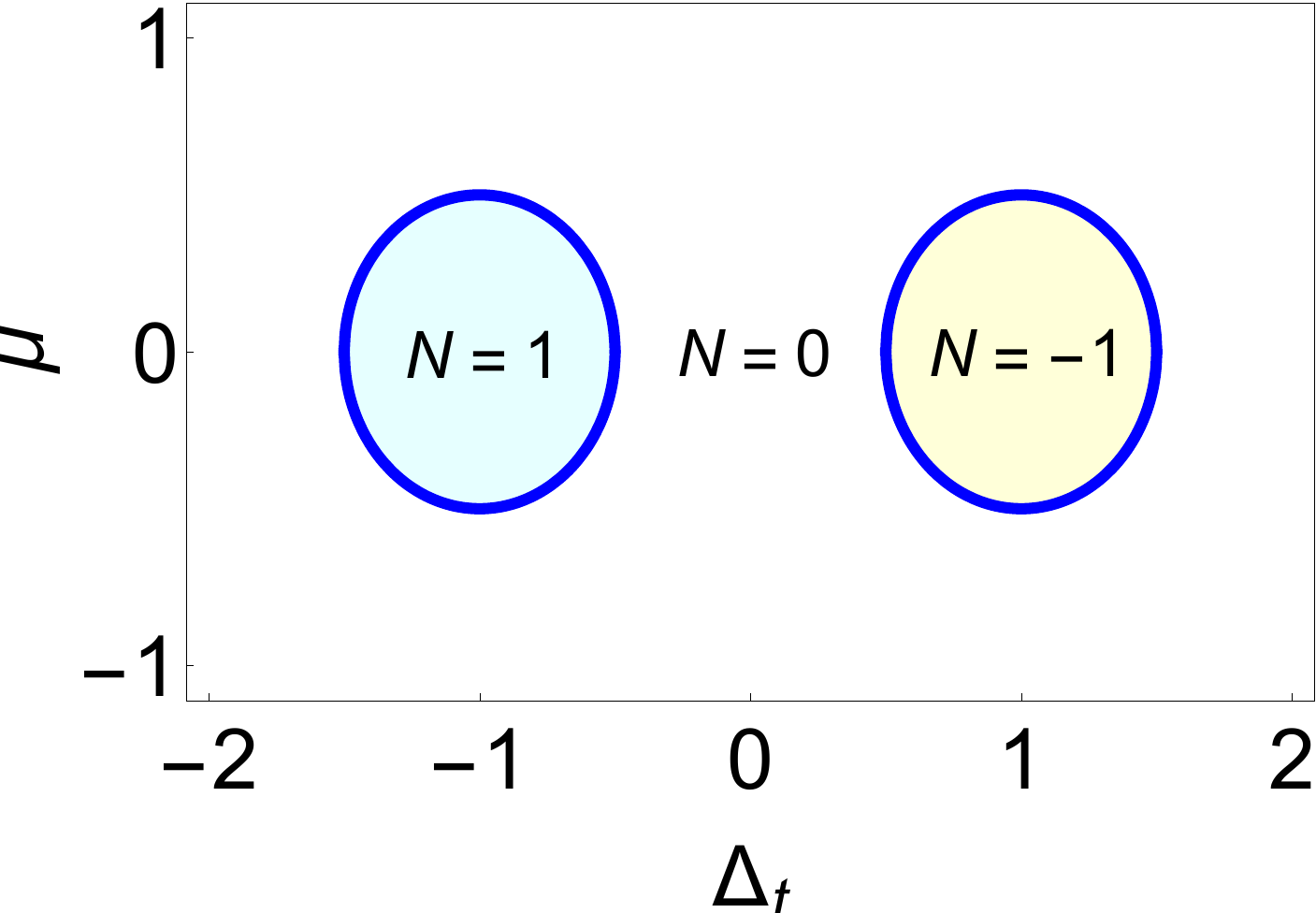}
	}   
\end{minipage}
  	\begin{minipage}[b]{0.23\textwidth}
 			\centering
 			\subfigure[]{
 				\includegraphics[width=1\textwidth]{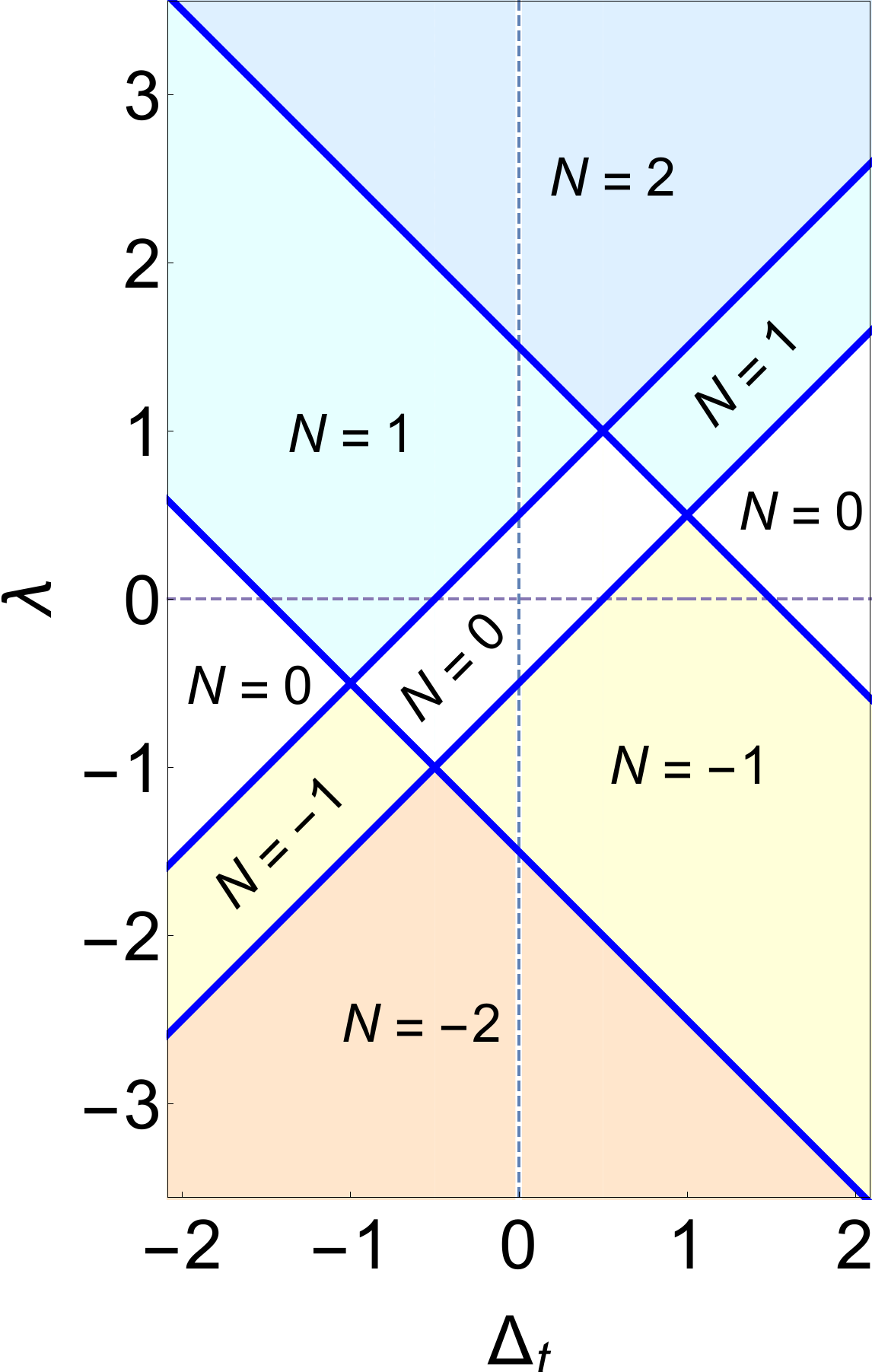}
 			}
 	\end{minipage}	
 	\caption{(color online)
 		Topological phase diagrams in (a) $\Delta_t$-$\Delta_s$ plane for $\mu=0$, (b) $\Delta_t-\mu$ plane for $\Delta_s=0$, and (c) $\Delta_t$-$\lambda$ plane for $\mu=0$ and $\Delta_s=0$.
 In (a)-(c), $t=0.5$, $N$ is the total Chern number, and blue lines are the phase boundaries.}
 		\label{fig3}
 \end{figure}

To determine the topological phase structure, we numerically calculate the Berry phase of each band. The topological property of the system is characterized by the total Chern number, i.e. the summation of the Berry phases of the lower four quasihole bands \cite{TKNN}. The regions marked by Chern number $N=\pm1$ on the topological phase diagrams in the $\Delta_t$-$\Delta_s$ plane at $\mu=0$ ({Fig.~\ref{fig3}a}) and in the $\Delta_t-\mu$ plane at $\Delta_s=0$ {(Fig.~\ref{fig3}b)}
correspond to the $\chi$TSC supporting a single $\chi$MEM at the boundaries.

The phase structures (Figs.~\ref{fig3}a and 3b) indicate that the $\chi$TSC is most prominent in the region of small $\mu$ and $\Delta_s$. It is therefore instructive to explore the nature of the topological phases and phase transitions in the limit of $\mu,\Delta_s=0$. In this case, the solution of the BdG Hamiltonian $h$ in Eq.~(\ref{h}) can be obtained analytically, even in the presence of an additional TR symmetry breaking Zeeman coupling
$h^\lambda=\lambda\sigma_z\chi_0\tau_0$, 
giving rise to $8$ quasiparticle/quasihole bands {
\begin{equation}
E_{\mu,\Delta_s=0}=\pm\sqrt{M_a^2+q^2},
\end{equation}
where} $a=1,...,4$ and {$M_1=\Delta_t-\lambda+\Delta+ t$, $M_2=\Delta_t+\lambda+\Delta- t$, $M_3=\Delta_t+\lambda-\Delta+ t$, and $M_4=\Delta_t-\lambda-\Delta- t$}.
Thus, there is an emergent Lorentz invariance in this limit, as all the BdG bands become relativistic with the corresponding effective mass $M_a$.
The Chern numbers associated with the four quasihole bands are,
$\frac{1}{2}{\rm sgn}(M_1)$, $-\frac{1}{2}{\rm sgn}(M_2)$, $-\frac{1}{2}{\rm sgn}(M_3)$, and $\frac{1}{2}{\rm sgn}(M_4)$, respectively. Whenever there is a sign change in $M_a$, i.e., a band inversion, the total Chern number changes by one. These sign changes as the Zeeman coupling $\lambda$ and {$\Delta_t$} are varied determine the topological phase diagram in the {$\Delta_t$-$\lambda$} plane (Fig.~\ref{fig3}c). The phases with the total Chern number $N=\pm1$ are indeed the $p$-wave $\chi$TSC \cite{RG}. To see this, consider, e.g. the phase with $\lambda=0$, $t>\Delta>0$, and {$-\Delta-t<\Delta_t<-t$}. The effective BdG Hamiltonian for the two bands closest to zero energy is given by
\begin{equation}
h=\left(
\begin{array}{cccc}
-M_1  & q_x-iq_y \\
q_x+iq_y & M_1
\end{array}
\right), \label{p-wave}
\end{equation}
which is identical to that of a $q_x+iq_y$-wave  $\chi$TCS with an effective chemical potential $M_1>0$.

Note that the phases along the $\Delta_t=0$ line in Fig.~\ref{fig3}c are consistent with those proposed in the quantum anomalous Hall insulator (QAHI)-SC proximity hybrid structures  {\cite{zhang3,science,hih}}, where ferromagnetism (thus $\lambda$) is crucial for breaking the TR symmetry. In contrast, along the $\lambda=0$ line in Fig.~\ref{fig3}c, the TR symmetry is broken by the inter-surface spin-triplet pairing of the SC-TSS, which is nonmagnetic and produces the $N=\pm1$ \emph{intrinsic} $p$-wave $\chi$TSC  with a single $\chi$MEM. The half-integer quantized conductance $\frac{e^2}{2h}$ of the $\chi$MEM can be detected directly at the edge of an antidot, as discussed in the SI \cite{SI}.

\begin{figure}	
\includegraphics[width=3.4in]{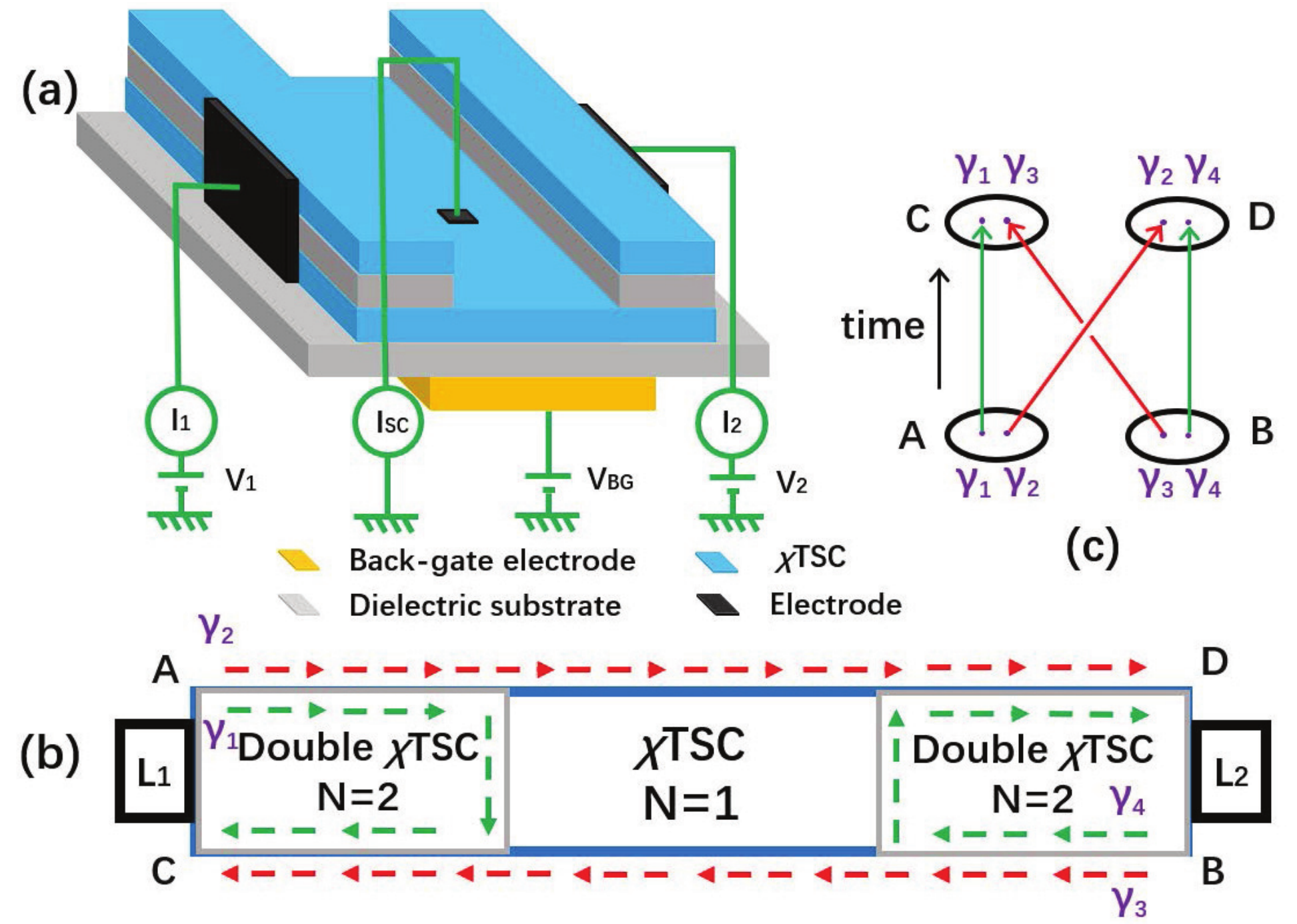}
	\caption{(color online)
		(a) A schematic of a layered $\chi$TSC structure for nonabelian braiding. The $N=2$ phase is achieved by the $\chi$TSC-insulator-$\chi$TSC structure at the two ends. (b) The top view of the device in (a). The arrows stand for the {$\chi$MEMs} which are labeled by $\gamma_i$.  (c) The electron at the leads splits into two {$\chi$MEMs} and becomes a nonlocal fermion. Under the braiding of $\gamma_2$ and $\gamma_3$, the evolution of the odd parity electron state from terminals A and B to C and D is equivalent to a Hadamard gate followed by a Pauli-Z gate.\label{fig6}}
\end{figure}

We turn to TQC using the $\chi$MEMs in such $\chi$TSCs. A protocol for nonabelian braiding based on the $\chi$MEMs has been proposed using the QAHI-(SC+QAHI)-QAHI proximity structures \cite{PNAS}. In the QAHI, the chiral fermion edge state can be viewed as two $\chi$MEMs; one of them propagates to the other side through the edge of the hybrid TSC (SC+QAHI) in the middle region, while the other cannot. Therefore, ideally, the junction braids the $\chi$MEMs and forms a nonabelian gate, which is equivalent to a Hadamard gate followed by a Pauli-Z gate \cite{PNAS}. However, the nonabelian braiding in such a system may fail because the SC+QAHI hybrid structure may end up in a metallic phase instead of the desired TSC \cite{JW}. Furthermore, there can be vortices in the superconductor, trapping MZMs and change the results of braiding dramatically \cite{nonabelian1,nonabelian2}.
These obstacles can be avoided by using the $\chi$TSCs. In order to realize the 2D spinor braiding statistics, four Majorana fermions are needed, which form an $SO(4)$ nonabelian group for the statistics \cite{NW,Ivanov}. We therefore construct a novel device with a common thin film $\chi$TSC shared by two layered $\chi$TSC-insulator-$\chi$TSC mesas on the sides ({Fig.~\ref{fig6}a}). Each mesa has a Chern number $N=2$ and supports two $\chi$MEMs on the edges of the top and bottom $\chi$TSCs, respectively ({Fig.~\ref{fig6}b}). For an electron in the leads with energy within the SC gap, there are two channels for transport: to enter the bulk $\chi$TSC by Andreev reflection or to split into two $\chi$MEMs and propagate along the edges. Once Andreev reflection is suppressed, the only open channel is through the $\chi$MEMs ({Fig.~\ref{fig6}b}) by splitting the electron into a nonlocal fermion. This can be achieved via single electron tunneling controlled by a gate-induced barrier at the contacts between the leads and the $\chi$TSC.

In the low current limit, the incoming electron states at terminals A and B can be represented by the four spatially separated Majorana fermions ({Fig.~\ref{fig6}b}): $|n_A^{\gamma_1\gamma_2}n_B^{\gamma_3\gamma_4}\rangle$ where $n_{A,B}$ are the fermion parities. For example, $|1_A\rangle=\psi^\dag_A|0\rangle$ with $\psi_A=\gamma_1+i\gamma_2$. The outgoing {states} at terminals C and D are, similarly, given by $|n_C^{\gamma_1\gamma_3}n_D^{\gamma_2\gamma_4}\rangle$. The dimension of the state space for a fixed fermion parity is two, which corresponds to a qubit. Under a braiding operation $\gamma_2\rightarrow\gamma_3, \gamma_3\rightarrow-\gamma_2$, the evolution of the parity odd electron state is thus equivalent to a Hadamard- followed by a Pauli-Z gate ({Fig.~\ref{fig6}c})
		\begin{equation}
		\left(
		\begin{array}{ccc}
		|0_C^{\gamma_1\gamma_3}1_D^{\gamma_2\gamma_4}\rangle \\
		|1_C^{\gamma_1\gamma_3}0_D^{\gamma_2\gamma_4}\rangle
		\end{array}
		\right)=\frac{1}{\sqrt{2}}\left(
		\begin{array}{ccc}
		1  & 1 \\
		-1 &  1
		\end{array}
		\right)\left(
		\begin{array}{ccc}
		|0_A^{\gamma_1\gamma_2}1_B^{\gamma_3\gamma_4}\rangle \\
		|1_A^{\gamma_1\gamma_2}0_B^{\gamma_3\gamma_4}\rangle
		\end{array}
		\right).
		\end{equation}
\begin{figure}
	\includegraphics[width=0.45\textwidth]{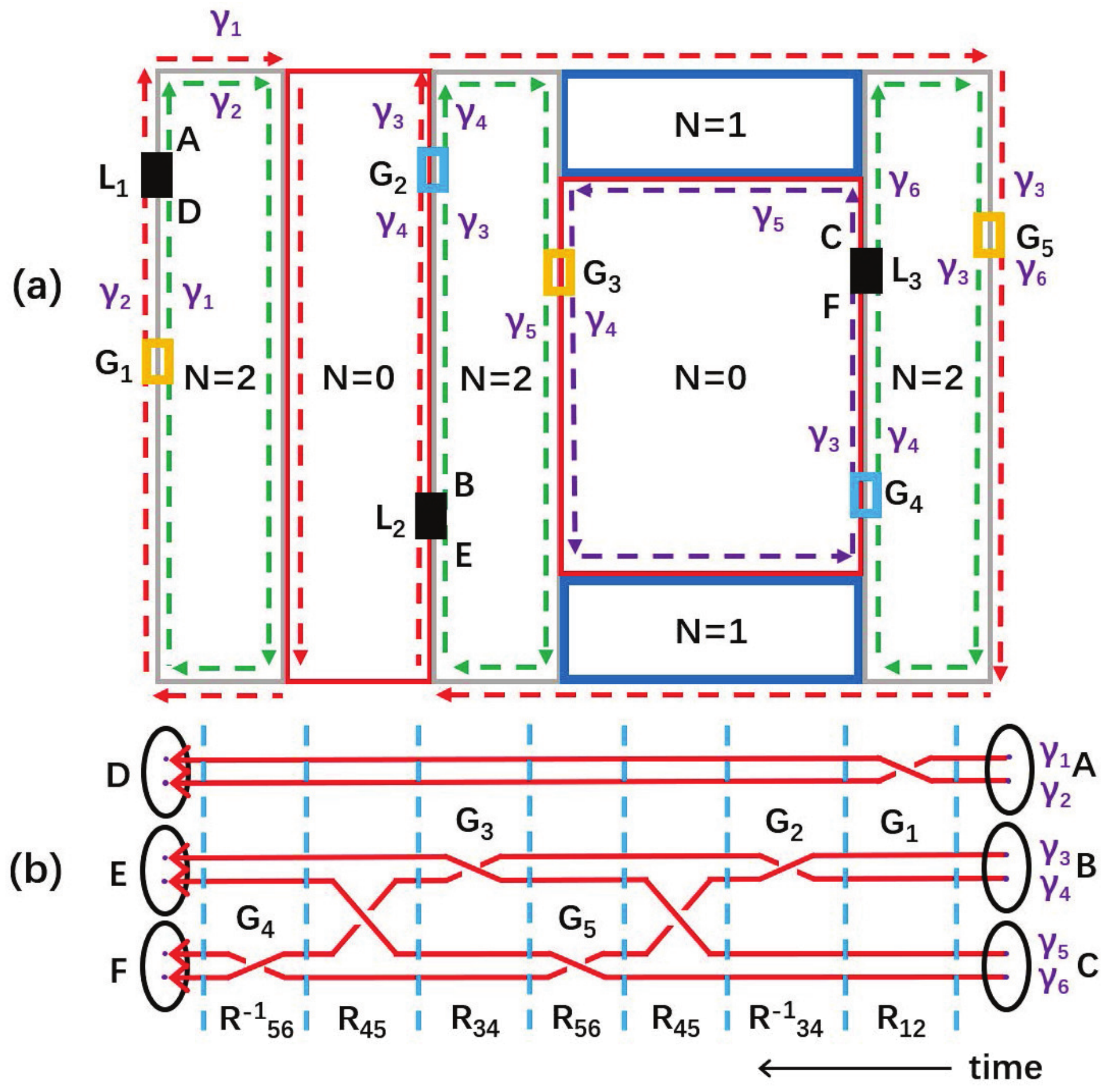}
	\caption{(color online)
		{(a) The top view of the device for the CNOT gate in Fig. \ref{figX}c. The substrate is a trivial insulator. {A} $\chi$TSC layer with $N=1$ covers the substrate except for the $N=0$ region. The $N=2$ region is the $\chi$TSC-insulator-$\chi$TSC mesa structure (see Fig.~\ref{figX}c and Fig.~\ref{fig6}). The solid black rectangles are the leads while the hollowed ones are the phase gates. For the CNOT gate in the even fermion parity sector, the gate-voltage induced phase shift is $\theta=\frac{\pi}{2}$ for $G_1$, $G_3$, and $G_5$, and $\theta=-\frac{\pi}{2}$ for $G_2$ and $G_4$. (b) The braiding diagram for the CNOT gate. The dashed blue lines stand for time slices.
The braiding $R$-matrices are explained in the text.}\label{fig6-2}}
\end{figure}

	To achieve universal quantum computing, we design a CNOT gate between 2-qubits using six $\chi$MEMs $\gamma_1,...,\gamma_6$ \cite{TQC1,TQC2}. The device is shown in Fig.~\ref{figX}c and its top view in Fig.~\ref{fig6-2}a.
{In the basis of even total fermion parity sector, i.e., $(|0_A0_B0_C\rangle$, $|0_A1_B1_C\rangle$, $|1_A0_B1_C\rangle$, $|1_A1_B0_C\rangle)^T$}, the counter-clockwise braiding matrices $R_{ij}$ between the $\gamma$'s are $R_{12}={\rm diag}(1,1,-i,-i)$, $R_{34}={\rm diag}(1,-i,1,-i)$, {$R_{56}={\rm diag}(1,-i,-i,1)$}, and
		\begin{eqnarray}
		R_{23}=
		\!\frac{1}{\sqrt{2}}\!\left(
		\begin{array}{cccc}
		1  & 0 & 0 & 1 \\
		0  & 1 & 1 & 0 \\
		0  & -1 & 1 & 0 \\
		-1 & 0 & 0 & 1
		\end{array}
		\right),
		R_{45}=\!
		\frac{1}{\sqrt{2}}\!\left(
		\begin{array}{cccc}
		1  & 1 & 0 & 0 \\
		-1  & 1 & 0 & 0 \\
		0  & 0 & 1 & 1 \\
		0 & 0 & -1 & 1
		\end{array}
		\right) \nonumber
		\end{eqnarray}
where the subscripts stand for the relative positions of the $\gamma$'s at a given time slice (dashed blue lines in Fig.~\ref{fig6-2}b). $R_{12}$, $R_{34}$, and $R_{56}$ can be realized by tuning the corresponding gate voltages $V$ over a length $L$, e.g., {$G^{ij}(\theta=VL)|1^{\gamma_i\gamma_j}\rangle=\exp (-i\theta)|1^{\gamma_i\gamma_j}\rangle$, and $G^{ij}(\theta)|0^{\gamma_i\gamma_j}\rangle
=|0^{\gamma_i\gamma_j}\rangle$. We will omit the upper indices when there is no confusion. Thus, $R_{12}={\rm diag}(1,1,G(\frac{\pi}{2}),G(\frac{\pi}{2}))$,} etc. On the other hand, $R_{23}$ and $R_{45}$ are naturally achieved by the delocalization of the two $\chi$MEMs at the edges of the two $N=1$ $\chi$TSCs which are topologically protected (Fig.~\ref{fig6-2}). The CNOT gate can therefore be achieved by a proper sequence of the $R$-matrices (Fig.~\ref{fig6-2}b)
\begin{equation}
		{\rm CNOT}=R_{56}^{-1}R_{45}R_{34}R_{56}R_{45}R^{-1}_{34}R_{12}
		=\left(
		\begin{array}{cccc}
		1  & 0 & 0 & 0 \\
		0  & 1 & 0 & 0 \\
		0  & 0 & 0 & 1 \\
		0  & 0 & 1 & 0
		\end{array}
		\right), \nonumber
\end{equation}	
where $R_{12}$ commutes with all other matrices.
In designing the quantum gates, the operators associated with the propagation of the $\chi$MEMs are topological. Although the ones corresponding to the gate phase shift $\theta$ are nontopological, the universal quantum computing {based} on the Ising type $\chi$MEMs is highly fault-tolerant \cite{TQC2,TQC3}. In other words, with the designed single-, double-qubits  and the supplementary phase gates,  universal quantum computing can be performed as long as the $\chi$MEMs are coherent. The coherence of the $\chi$MEMs can be tested through the oscillation in the conductance $\sigma_{23}$ between leads 2 and 3 in Fig.~\ref{fig6-2}(a) by continuously tuning one of the phase gates between {$G_{3}$ and $G_{5}$} while keeping all other gates stationary, similar to that in the Corbino structure discussed in Ref. \cite{PNAS}.

In summary, we have shown that $\chi$TSCs can emerge from coupling the SC-TSS in superconductors with a topologically nontrivial band structure, such as in the Fe-based superconductor FeTeSe. The spin-triplet, inversion odd inter-surface pairing breaks time-reversal symmetry without magnetism and produces a 2D $\chi$TSC with a single $\chi$MEM at the boundary and MZM in the vortex core. The breaking of inversion symmetry under interchanges the two surfaces by gating or different substrates generally leads to mixing of singlet and triplet inter-surface pairing and increases the potential for realizing such intrinsic TSCs. Nonabelian braiding of the $\chi$MEMs and the detection of the half-integer quantized Hall conductance (see SI \cite{SI}) using the $\chi$TSC based devices can be more advantageous over the earlier proposal using the QAHI and SC proximity structures. Such devices can be fabricated using thin films of Fe-based superconductors with SC-TSS (Fig.~\ref{figX}a). We have proposed a novel CNOT gate device for universal quantum computing using such $\chi$TSC. A different double-surface structure is the opposing surfaces of two superconductors supporting the SC-TSS (Fig.~\ref{figX}b), e.g., the Josephson junction of two Fe-based superconductors. To explore the Majorana physics in such bilayer structures is a future task.

%\noindent{\bf Acknowledgements}
{We thank Bin Chen, Dong-Lai Feng,  Zheng-Cheng Gu, Kun Jiang, Jing Wang, Yi-Hua Wang, Yong-Shi Wu, and Sen Zhou for helpful discussions. This work is supported by NNSF of China with No. 11474061 (XL,YGC,YY), No. 11804223 (XL) and the U.S. Department of Energy, Basic Energy Sciences Grant No. DE-FG02-99ER45747 (ZW).}

%\noindent{\bf Contributions}

\end{document}